\documentclass[prb,preprint]{revtex4} 

\usepackage{amsmath} 

\usepackage{graphicx} 

\newcommand{\be}{\begin{equation}}
\newcommand{\ee}{\end{equation}}
\newcommand{\bea}{\begin{eqnarray}}
\newcommand{\eea}{\end{eqnarray}}

\begin{document}

\title{The gravitational field of a cube}
\author{James M.~Chappell}
\email{james.m.chappell@adelaide.edu.au}
\affiliation{School of Electrical and Electronic Engineering, University of Adelaide, South Australia
5005, Australia}
\author{Mark Chappell}
\affiliation{Griffith Institute, Griffith University, Queensland 4122, Australia}
\author{Azhar Iqbal}
\affiliation{School of Electrical and Electronic Engineering, University of Adelaide,
South Australia 5005, Australia}
\author{Derek Abbott}
\affiliation{School of Electrical and Electronic Engineering, University of Adelaide,
South Australia 5005, Australia}
\date{\today}

\begin{abstract}
Large astronomical objects such as stars or planets, produce approximately spherical shapes due to the large gravitational forces, and if the object is rotating rapidly, it becomes an oblate spheroid.  In juxtaposition to this, we conduct a thought experiment regarding the properties of a planet being in the form of a perfect cube.  We firstly calculate the gravitational potential and from the equipotentials, we deduce the shape of the lakes that would form on the surface of such an object. We then consider the formation of orbits around such objects both with a static and a rotating cube.  
A possible practical application of these results is that, because cuboid objects can be easily stacked together, we can calculate the field of more complicated shapes, using the principle of superposition, by simply adding the field from a set of component shapes \cite{Mufti1975}.
\end{abstract}

\maketitle

\section{Introduction}
Plato, in the Timaeus, links what he considers the four basic elements, fire, air, water and earth, with the four regular solids, using the tetrahedron to represent fire, the octahedron to represent air, the icosahedron to represent water and the cube used to represent the earth.
Taking Plato's idea literally, we might imagine the earth in the form of a perfect cube.  We can then calculate the gravitational field around this object, the process being simplified by the simple endpoints needed when integrating over a cuboid mass, assuming Newton's classical law of gravitation.  The equipotentials of the field will give us, for example, the edge of a lake that would form on the face of such a cube.  To make the problem tractable, we will ignore the mass of the fluid, and assume there is zero surface tension.  We also investigate the nature of the satellite orbits that would form around such an object.  The gravitational field of a cube was first calculated in 1958 \cite{MACMILLAN}, and later with refinements \cite{Nagy_1966,Mufti}, and also with varying density \cite{Abdeslem}, although with a fictitious cubic planet, a uniform density would seem appropriate as a first approximation.

\section{Analysis}

\subsection{Gravitational potential of a cube}

We first seek the Newtonian gravitational potential of a rectangular solid of uniform density~$ \rho $ with Newton's universal gravitational constant $G$.
We suppose the rectangular solid has a length $ L $, breadth $ B $ and depth $ D $, centered on the origin, then we have the potential
\bea
V(X,Y,Z) & = & - G \rho \int_{-D}^D \int_{-B}^B \int_{-L}^L \frac{ d x' d y' d z' }{\sqrt{(X-x')^2+(Y-y')^2+ (Z-z')^2}} \\ \nonumber
 & = & - G \rho \int_{z=-D-Z}^{D-Z} \int_{y=-B-Y}^{B-Y} \int_{x=-L-X}^{L-X} \frac{ d x d y d z }{\sqrt{x^2+y^2+ x^2}} \\ \nonumber
 & = & - G \rho \int_{x=-L-X}^{L-X} \int_{y=-B-Y}^{B-Y} \left [ \ln(z+\sqrt{x^2+y^2+z^2}) \right ]_{z=-D-Z}^{D-Z} d y d x , \nonumber
\eea
where we made the substitution, $ x = x' - X $, $ y = y' - Y $ and $ z = z' - Z $, and completed the integral over the $ z $ coordinate.
Next, integrating over the $ y $ variable, and using $ r =  \sqrt{x^2+y^2+z^2} $, we find
\bea \label{potentialCubezyInegrals}
V & = & -G \rho \int_{x=-L-X}^{L-X} d x \Big [ \Big [ y \ln \left (z+r \right ) + z \ln \left (y+r \right )  \\ \nonumber
& & - y + x \arctan \frac{y}{x} - x \arctan \frac{yz}{x r} \Big ]_{z=-D-Z}^{D-Z}  \Big ]_{y=-B-Y}^{B-Y}  \nonumber
\eea
and with the integral over $ x $, we achieve our final result
\bea \label{potentialCube}
V(X,Y,Z) & = & -G \rho \Big [ \Big [ \Big [  y z \ln \left (x+r \right ) - \frac{x^2}{2} \arctan \frac{y z}{x r}  + x z \ln \left (y+r \right )  - \frac{y^2}{2} \arctan \frac{x z}{y r}  \\ \nonumber
& & + x y \ln \left (z+r \right )  - \frac{z^2}{2} \arctan \frac{x y}{z r}  \Big ]_{z=-D-Z}^{D-Z}  \Big ]_{y=-B-Y}^{B-Y}  \Big ]_{x=-L-X}^{L-X}  \nonumber
\eea
or using $ x_1 = x, x_2 = y, x_3=z $, and $ D_1 = L, D_2 = B, D_3 = D$ we can write
\be
V(X_1,X_2,X_3)  = -G \rho \sum_{j=1}^3  \Big [ \sum_{i=1}^3  \Big ( \frac{v}{x_i} \ln \left (x_i+r \right ) - \frac{x_i^2}{2} \arctan \frac{v}{x_i^2 r} \Big )  \Big ]_{x_j=-D_j-X_j}^{D_j-X_j}  
\ee
for the gravitational potential of a cuboid mass, where $ v = x_1 x_2 x_3 $.  The pairs of $ \log $ and $\arctan $ terms combine to produce the expected $ \frac{1}{r} $ falloff in gravitational potential at large distances. For example, for a $ 2 \times 2 \times 2 m^3 $ cube, with $ G \rho = 1 $,  as $ x \rightarrow \infty $, $  y z \ln \left (x+r \right ) - \frac{x^2}{2} \arctan \frac{y z}{x r} \rightarrow -\frac{8}{r} $, as expected for a point source.

\subsection{A lake on the face of a cube}

\begin{figure}[hbtp]  
\includegraphics[height=80mm]{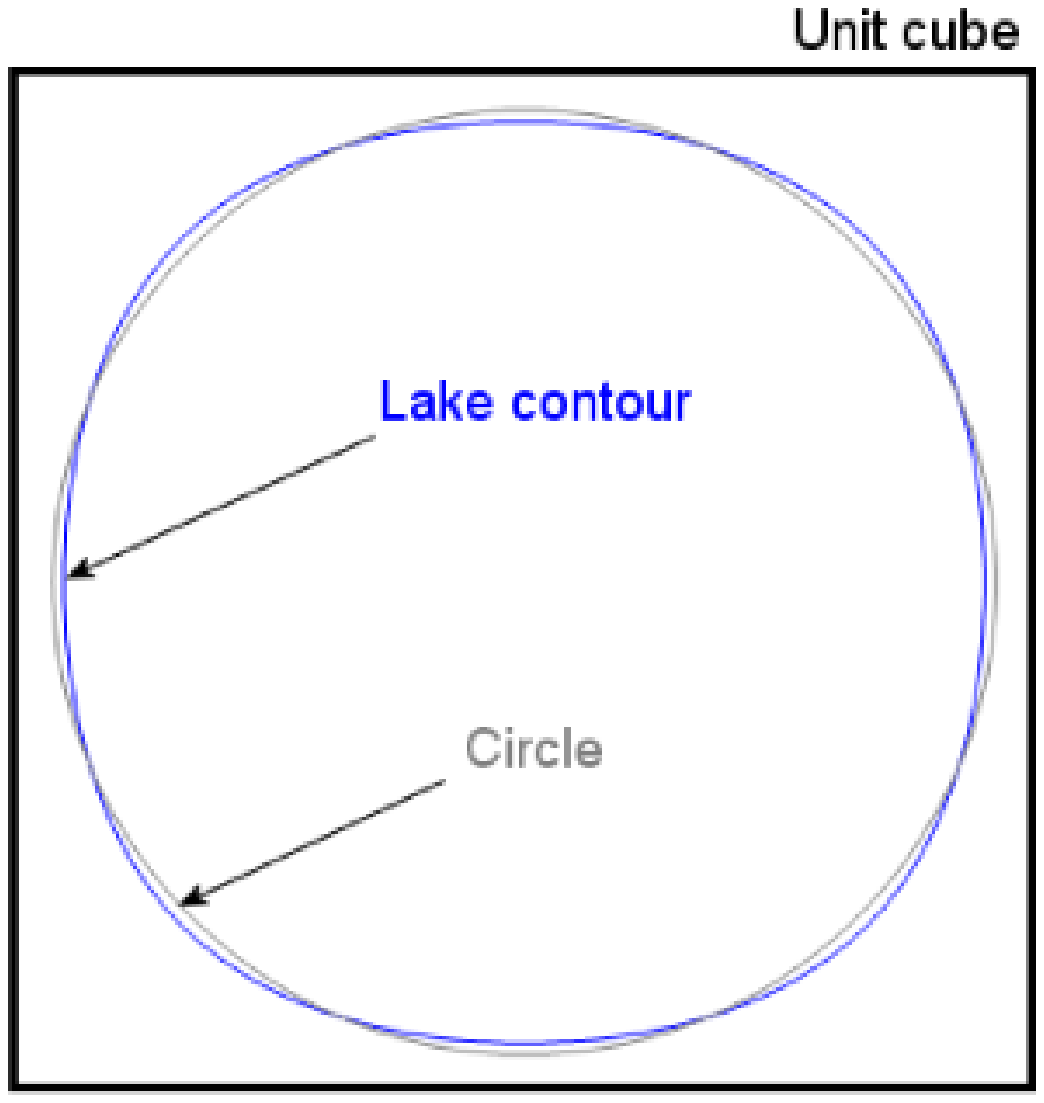}   
\includegraphics[height=80mm]{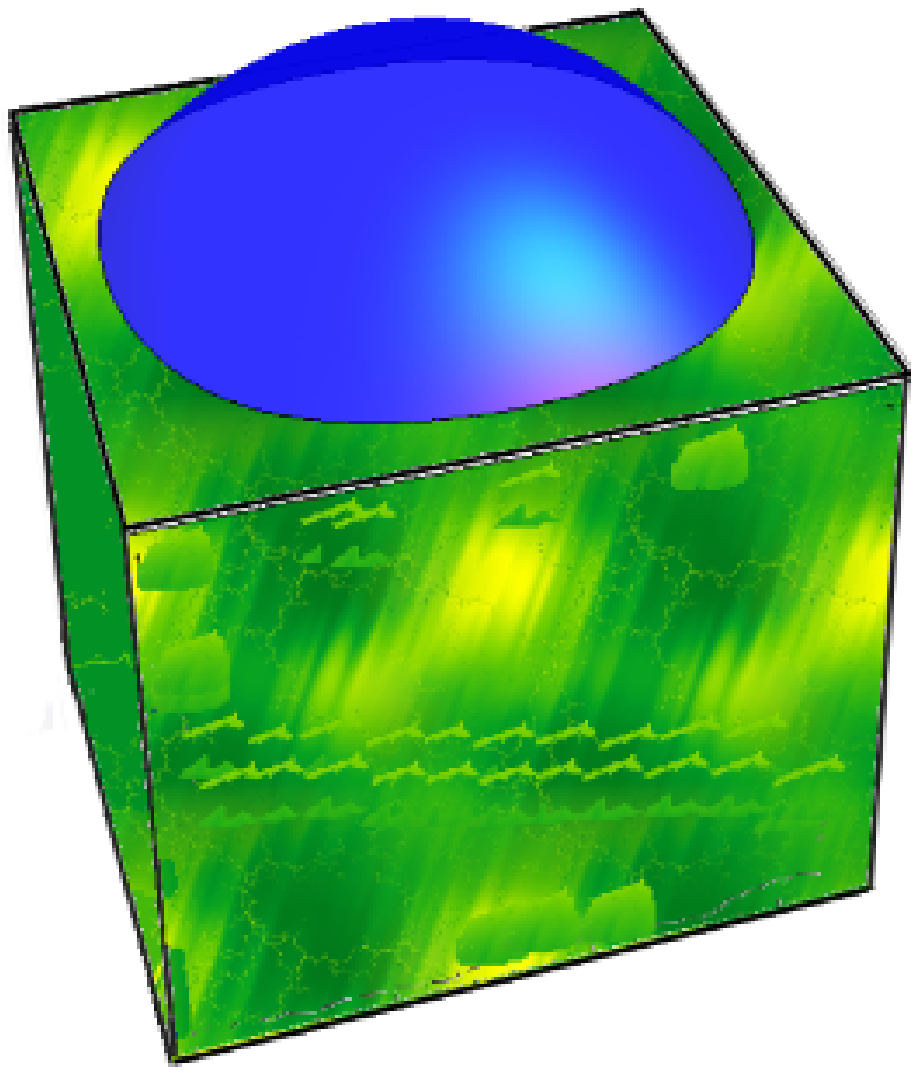} 
\caption{A lake formed on the surface of a cube. As expected, the edge of the lake is `pulled up' towards the corners, due to the extra mass present there, but forming a nearly spherical surface. \label{lakeCube}}
\end{figure}

In order to reveal the equipotentials around this object, we can imagine a very low density fluid being added to a face of a $ 2 \times 2 \times 2 $ cube.  We specify a very low density fluid in order not to modify the existing field, and by plotting Eq.~(\ref{potentialCube}), at a constant potential we find a lake as shown in Fig.~\ref{lakeCube}.  A perfect circle is shown for comparison, and we can see how the water is dragged up towards the corners.  

The surface of the lake, for someone boating on the surface, will be approximately spherical in shape, as it is for a large lake upon the earth, as is also shown in Fig.~\ref{lakeCube}, found by plotting Eq.~(\ref{potentialCube}) at a constant potential in three dimensions.

If we keep adding the fluid until the edge of the cube is reached, we would have a depth of $ 0.3346 $ units over the face, compared to $ \sqrt{2} -1 = 0.4142 $ for a spherical surface.
If we continue to add fluid until the corners are covered, we would then have a depth of $ 0.6389 $ units, above the center of the face, compared to $ \sqrt{3} -1 = 0.7321 $ for a spherical surface.

\subsection{Gravitational field}

We calculate the gravitational field vectors from
\be
\vec{g} =- \nabla V = -\left (\frac{\partial V }{\partial x } e_1 + \frac{\partial V }{\partial y } e_2 + \frac{\partial V }{\partial z } e_3 \right ) .
\ee
The field vector in the $ x $ direction $ g_x = -\frac{\partial V }{\partial x } $, can be deduced from Eq.~(\ref{potentialCubezyInegrals}), before the last integral is calculated, and hence by the fundamental theorem of calculus we find
\be \label{FieldEx}
g_X  =  G \rho \sum_{j=1}^3 \Big [ y \ln \left (z+r \right ) + z \ln \left (y+r \right ) - x \arctan \frac{yz}{x r} \Big ]_{x_j=-D_j-X_j}^{D_j-X_j}   
\ee
and from symmetry we can also easily deduce the field strengths in the $ y $ and $ z $ directions, giving the field strength vector $ \mathbf{g} = \sum_{i=1}^3 g_i e_i$, where
\be \label{FieldAcceleration}
g_i =  G \rho \sum_{b=1}^3 \Big [ x_j \ln \left (x_k+r \right )+x_k \ln \left (x_j+r \right ) - x_i \arctan \frac{v}{x_i^2 r} \Big ]_{x_b=-D_b-X_b}^{D_b-X_b} 
\ee
for distinct $ i,j,k $, where $ e_i $ are the unit vectors for the $ x,y,z$ coordinate system.

\begin{figure}[htb]

\begin{center}
\includegraphics[width=3.6in]{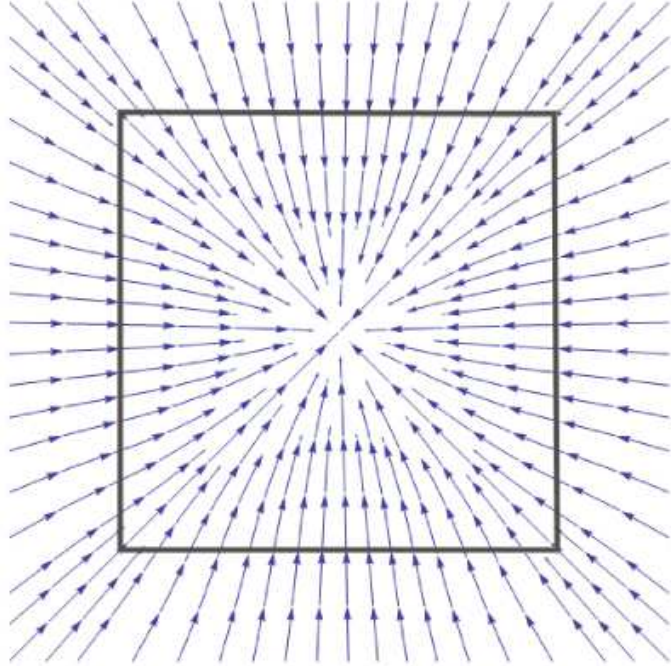}
\end{center}

\caption{The field through a cube sliced in half through the faces. We can observe the slight distortion of the field lines between the edges and the center of each face. \label{FieldLinesSlice}}
\end{figure}

If we look at the changing direction of the field as we move across a face, then we observe that the field vector only points towards the center of the cube at the center of each face, at the corners, and at the center of each edge, which could also be deduced by symmetry arguments, refer Fig.~\ref{FieldLinesSlice}.

\subsection{Orbits around the cube}

\begin{figure}[hbtp]  
\includegraphics[height=65mm]{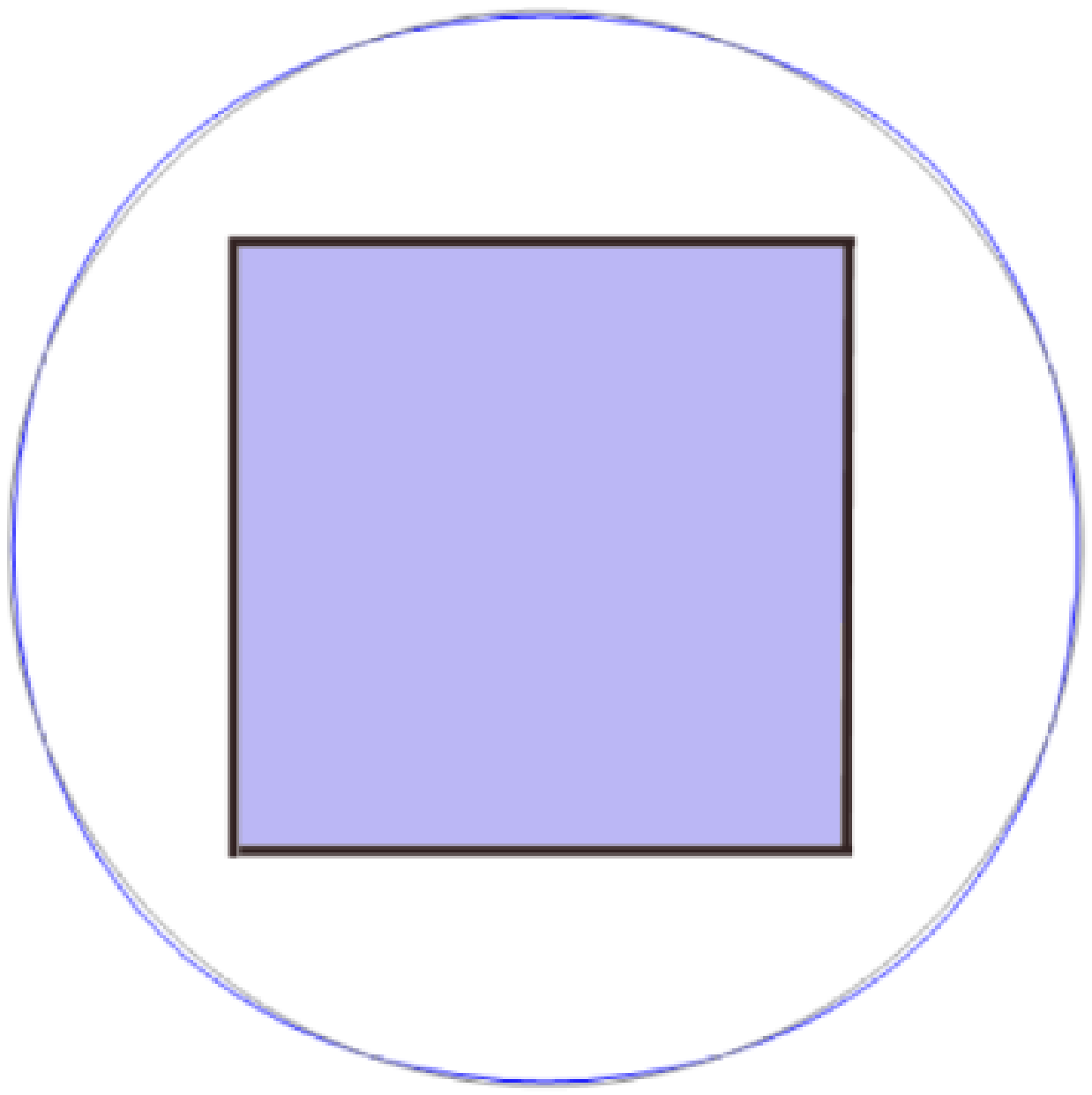}   
\includegraphics[height=55mm]{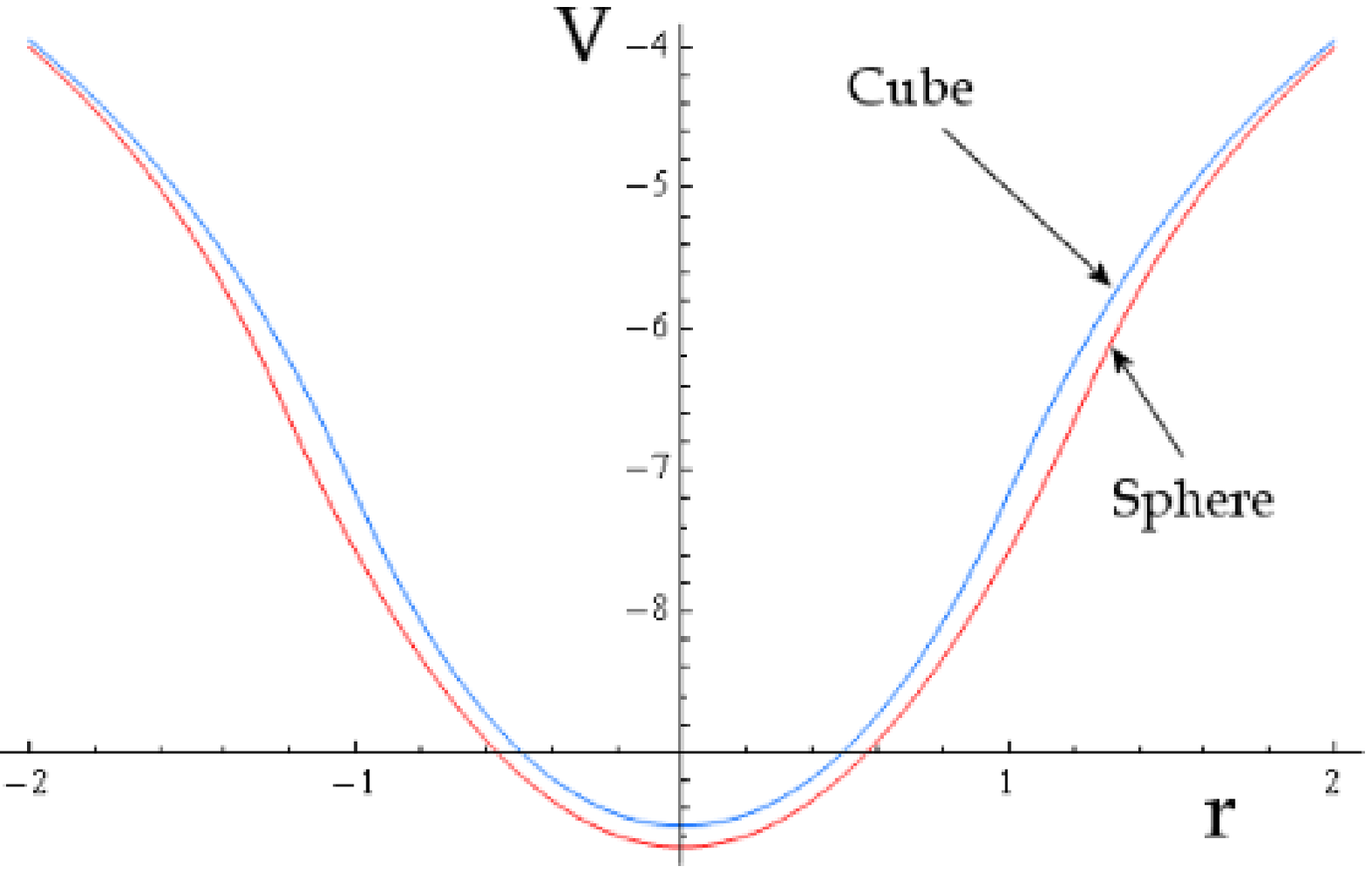} 
\caption{The equipotential around the equator of the cube.  We notice how the field is slightly stronger over the corners, indicated by the equipotential being shifted outwards compared to a perfect circle. When comparing the gravitational potential of a cube to a sphere of the same mass, because the sphere is a more compact object we find a deeper potential well, though the two potentials converge at larger distances as expected.\label{orbitCube}}
\end{figure}

Could a moon or satellite, orbit this cubic planet?  We notice that there is slightly greater gravitational force of attraction over the corners of the cube, and hence an orbiting satellite would significantly couple with the spin of the cube, refer Fig.~\ref{orbitCube}.

If we assume a satellite orbit around the faces and the centre of the edges we can reduce the orbit to the plane, and so need to solve the orbital equations
\be
\ddot{x} = g_x \, , \,\,\,  \ddot{y} = g_y ,
\ee
refer Eq.~(\ref{FieldAcceleration}).

\begin{figure}[htb]

\begin{center}
\includegraphics[width=4.2in]{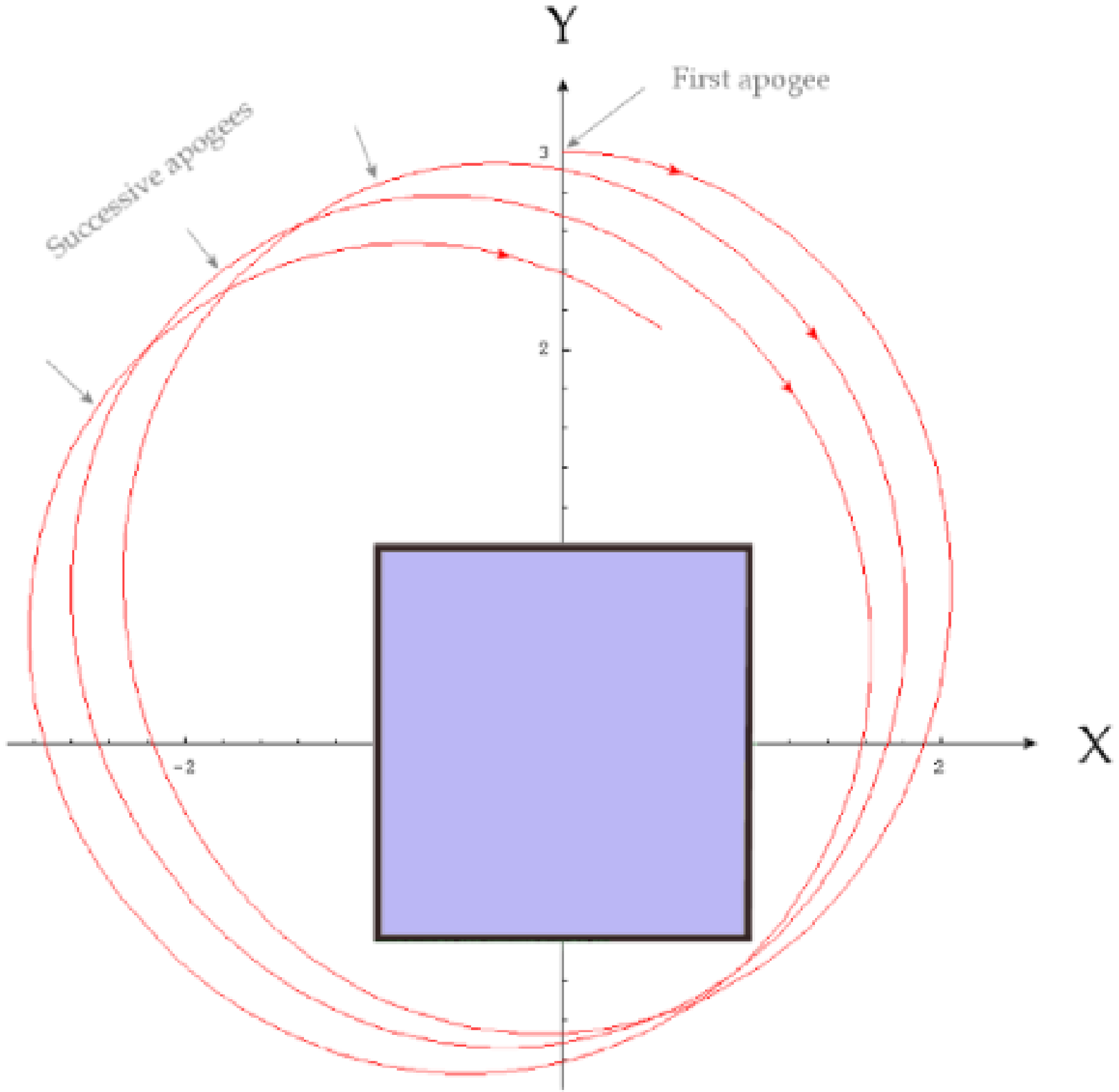}
\end{center}

\caption{A satellite orbiting around the equator of a cube with a side length equal to the diameter of the earth.  Beginning with an satellite orbital radius of three earth radii and a velocity of 3.63 km/s, we find a period of approximately 4.8 hours, but with an orbit that precesses fairly rapidly, as shown by the counterclockwise movement of successive apogees. \label{ExampleOrbitCube}}
\end{figure}

Solving this equation numerically for the specific case, of a satellite orbiting a cube with a side length equal to the diameter of the earth, with an initial height of three earth radii moving in the $ X $ direction as shown in Fig.~\ref{ExampleOrbitCube}, with a velocity of 3.63 km/s.  

We find an orbital period of approximately 4.8 hours, and due to the interaction over the corners, the orbit is distorted from a perfect ellipse, creating a fairly rapid precession as shown by the counterclockwise precession of the successive apogees in Fig.~\ref{ExampleOrbitCube}. The orbit is not closed, but provided the precession is a rational fraction of the orbit, the orbit will eventually close. 

Bertrand's theorem from classical mechanics states that `The only central forces that result in closed orbits for all bound particles are the inverse-square law and Hooke's law.' \cite{goldstein2002}  Due to the presence of a cuboid mass distribution the force field felt by the satellite is not inverse square and so we would not expect closed orbits.  

\begin{figure}[htb]

\begin{center}
\includegraphics[width=4.2in]{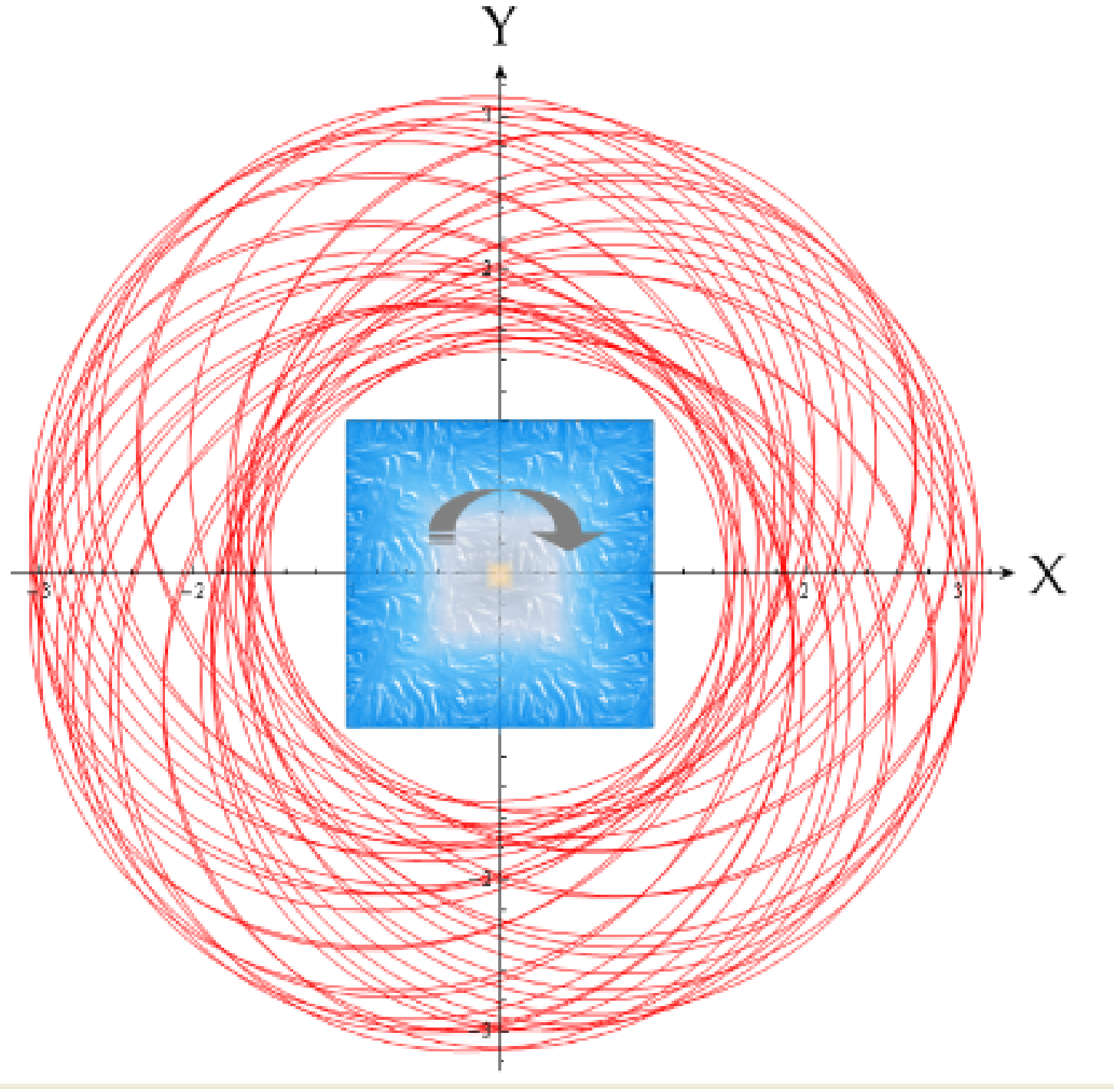}
\end{center}

\caption{Following from the previous figure except that the central cube is now rotating with a 24 hour day, in conjunction with the 4.8 hour satellite orbit. Both rotations are clockwise, but the irregular perturbation of the satellite by the cube creating a continually varying trajectory. \label{ExampleLongtime4OrbitCube}}
\end{figure}

\begin{figure}[hbtp]  
\includegraphics[height=77mm]{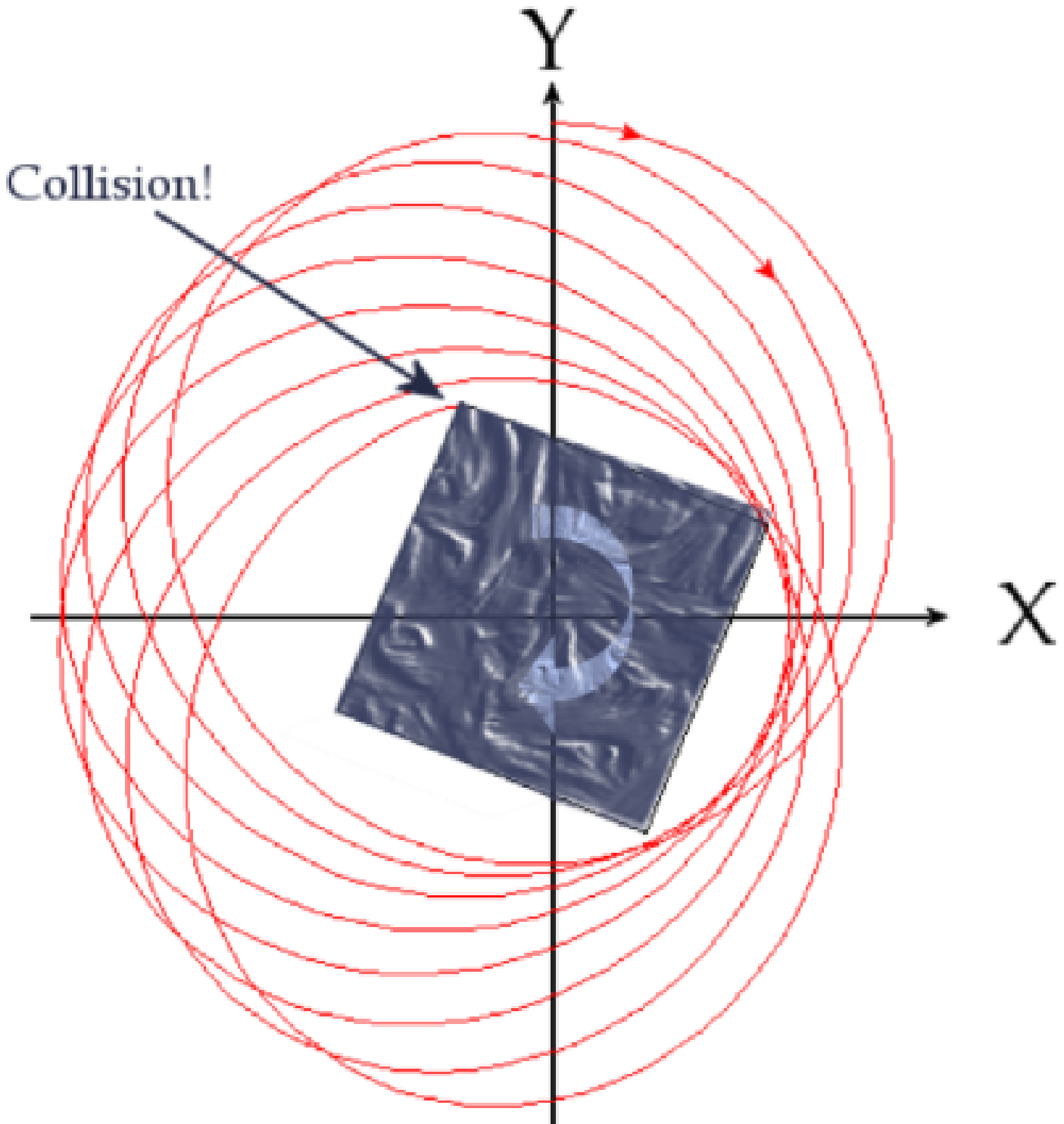}   
\includegraphics[height=67mm]{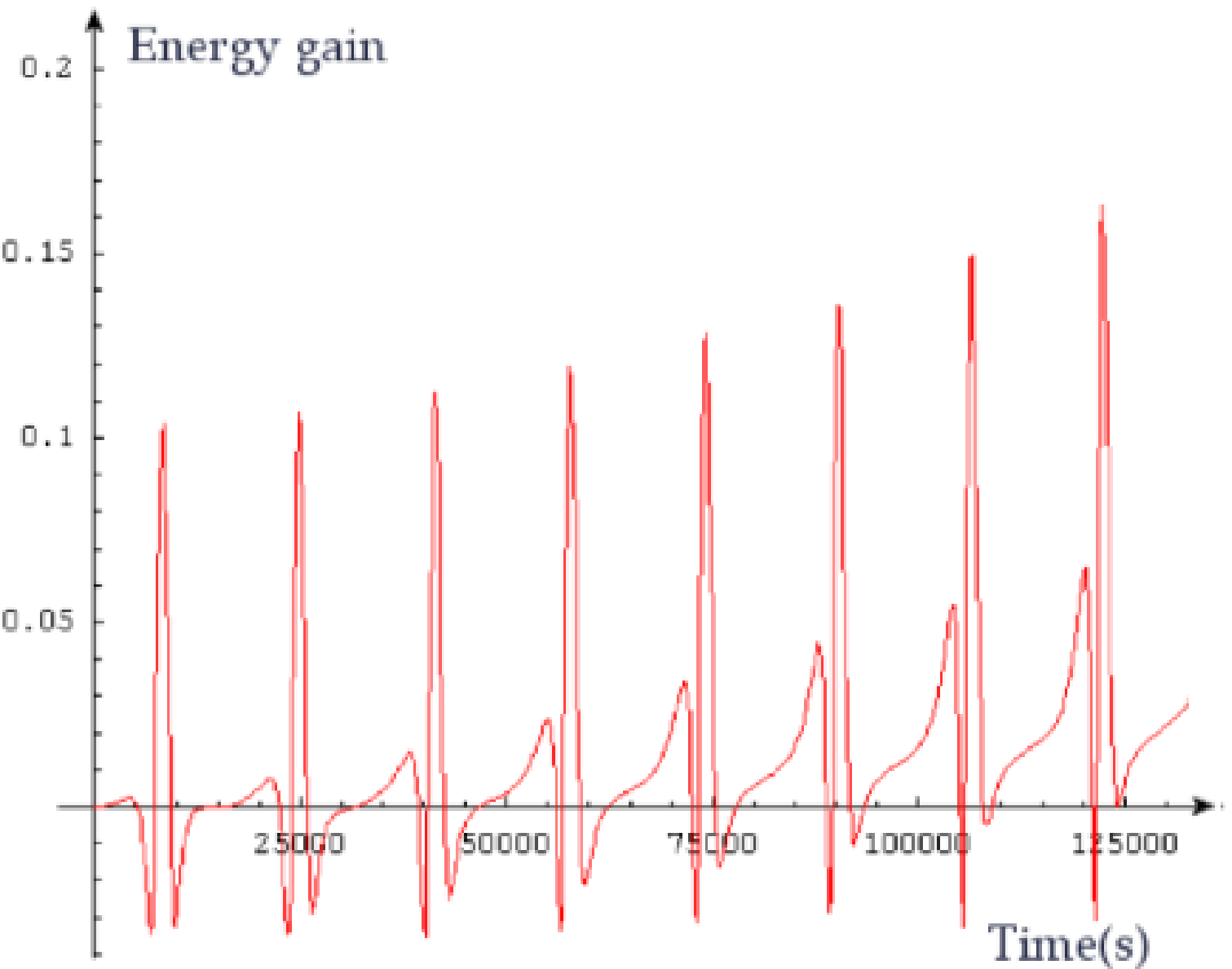} 
\caption{Following from the previous figure except that the cube is now rotating faster with a 10 hour day, in order to increase the resonance with the 4.8 hour satellite orbit. We can see that with this resonance the satellite rapidly acquires energy, colliding with the cube near the end of the 8th orbit as shown. From the graph of energy gain, we can see that by the 8th orbit the satellite gains nearly 17\% more energy at perigee. \label{ExampleCollision}}
\end{figure}

If we now investigate a satellite orbiting a rotating cube in the same direction as the cube rotation we find quite irregular orbits as shown in Fig.~\ref{ExampleLongtime4OrbitCube}. This must be due to the significant perturbation over the corners of the cube when the satellite is near perigee, which deflect the satellites orbit, but at irregular intervals due to the differential rotation periods. 
If we reduce the cube rotation to a 10 hour day, so that it is nearly exactly double the satellite period of 4.8 hours, in order to accentuate any resonance effects, then we find in the 8th orbit that the satellite now collides with the face of the cube, as shown in Fig.~\ref{ExampleCollision}.

The satellite is in fact picking up energy, analogous to the well known slingshot effect, as can be seen in Fig.~\ref{ExampleCollision}.
This energy is being extracted from the cube, however as the satellite periodically goes out of resonance, the energy will be returned to the cube.  Of course the energy of the combined system is unchanging.  It is not easily visible on Fig.~Fig.~\ref{ExampleCollision} but the apogee of the satellite is in fact increasing by a few percent that reflects the energy gain.

\section{Discussion}

Calculating the gravitational potential and vector field for a cubic mass, we find the shape of a lake which would form on the surface of such a cubic mass, and the overall shape of its spherical surface, shown in Fig~\ref{lakeCube}. 

We then examined the possibility of stable orbits around such an object. Solving orbital equations numerically we find that the orbits are not closed in general but precessed fairly rapidly due to interactions of the satellites with the mass in the corners of the cube. Orbits around a rotating cube are then investigated, which allow resonance effects between the cube rotation and the satellite orbital period, which show a significant effect on the orbit causing it to now crash into the cube, as shown in Fig.~\ref{ExampleCollision}.  

As an extension to our classical analysis we might also consider the calculation of the energy levels determined from the Schr$\ddot{\rm{o}}$dinger equation for the cube potential $ V $ given by $ \left (\frac{-\hbar^2}{2 m} \nabla^2 + V \right ) | \psi \rangle = E | \psi \rangle $.  We could also investigate whether an analytic solution is possible for a cuboid mass using the equations of general relativity.

For the dimensions of the cube being $ 12,000 \times 12,000 \times 12,000 \, \rm{km}^3 $, approximately the size of the earth, with the same volume of water and atmosphere as found on the earth, then we would approximately half fill each face with water and have an atmosphere approximately 20 km thick.
In this case then the corners and the edges of the cube, would be like vast mountain ranges several thousand km high, with their tips extending out into free space.  It would therefore be very difficult to cross these mountain ranges, and hence we would have six nearly independent habitable zones on each face.  There would  presumably be permanent snow on the sides of these vast mountain ranges and people would live around the edges of the oceans on each face in a fairly narrow habitable zone, which would be rather like living on the side of a fairly steep mountain bordering the ocean.  Unfortunately climbing the $ 3000$ km high corners does not result in an improved view because the surface is still flat in any observed direction. However the corners, being in free space, would be very suitable for launching satellites.  One would also have a 1000 km of downhill ski run from each corner, and one could pick which of three habitable zones one wanted to ski into.  
In order to create a day night cycle we would also need the cube to be rotating.
The sun would rise almost instantaneously over a cube however, so that the whole face would need to be a single timezone, and thus the cube as a whole would require four separate timezones.  The north and south faces in this case would be permanently frozen as they would receive no sunlight except that striking the oceans extending away from the surface of the cube, so there might be a permanent pool of liquid water at the two poles.

Launching low orbit satellites around this cube needs special care in order to avoid certain orbital resonances that would create significant variations in the orbit.

\appendix*


\end{document}